# Methods for Accelerating Conway's Doomsday Algorithm (part 2)

Chamberlain Fong [spectralfft@yahoo.com]
Michael K. Walters [mikewalters72@gmail.com]

**Abstract**
We propose a modification of a key component in the Doomsday Algorithm for calculating the day of the week of any given date. In particular, we propose to replace the calculation of the required expression:

$$\left\lfloor \frac{x}{12} \right\rfloor + x \, mod \, 12 + \left\lfloor \frac{x \, mod \, 12}{4} \right\rfloor$$

with

$$-\left[ \frac{x + 11(x \, mod \, 2)}{2} + 11\left(\frac{x + 11(x \, mod \, 2)}{2} \, mod \, 2\right) \right] (mod 7)$$

for a 2-digit input year *x*;

Although our expression looks daunting and complicated, we will explain why it is actually easy to calculate mentally.

## Introduction

The Doomsday algorithm is an algorithm for determining the day of the week for any calendar date. It is simple enough that, with practice, it can be done mentally without paper or pencil. Just like in part 1 of this paper, we would like to suggest a modification to the calculation of the doomsyear component of the Doomsday algorithm. We call this proposed method the *Odd+11 method*. For comparison, we will call the method proposed in part 1 of this paper as the *Decade method*. Both methods attempt to simplify the mental calculation[1] of Conway's doomsyear term:

$$\left\lfloor \frac{x}{12} \right\rfloor + x \, mod \, 12 + \left\lfloor \frac{x \, mod \, 12}{4} \right\rfloor$$

But, they approach the problem in very different ways. Indeed, the Odd+11 method was discovered during discussions of the Decade method. We could have written a completely separate paper to introduce the Odd+11 method, but we feel that both methods tackle the same problem and belong in a paper together; albeit, in different parts.

**The Odd+11 Method**

The Odd+11 method is an improved variant of Mike Walter's method originally proposed in 2008[2]. We provide a summary and proof of Mike Walter's method in the appendix of this paper. The Odd+11 method was designed with two criteria in mind: avoid temporary variables to remember while doing the calculation; and avoid modulo 4 calculations. Indeed, this method does not involve any modulo 4 operations. Instead, there are 2 odd/even checks and conditional additions of 11. Here is an outline of the method:

1) If the input year x is an odd number, add 11 to it. Otherwise, do not add anything
2) Divide the result by 2
3) If the result is an odd number, add 11 to it. Otherwise, do not add anything
4) Calculate the modulo 7 of the result to get a remainder
5) Subtract this remainder from 7 to get the doomsyear of x.

The procedure mentioned above can be expressed in the form an equation:

$$doomsyear(x) = -\left[\frac{x + 11(x\,mod\,2)}{2} + 11\left(\frac{x + 11(x\,mod 2)}{2}\,mod\ 2\right)\right] (mod\,7)$$

Although this expression looks daunting and complicated, it is actually simple because of a common subexpression

$$\frac{x + 11(x\,mod\,2)}{2}$$

that only needs to be calculated once.

We would like to emphasize that this equation is probably the worst possible way to remember the algorithm. We only provided it for reference and use in a mathematical proof of the correctness of the method later in this paper. The algorithm is best remembered by using the flowchart shown on the next page

As we previously mentioned, one of our design criterion for this method was to avoid temporary variables to remember in your head aside from a running-sum accumulator while doing the mental calculation. Indeed, Conway's doomsyear formula and the Decade method require extra temporary variables to recall while doing the mental calculation. We feel that the need to memorize temporary numbers while calculating something else makes the process error-prone; and hence, avoided it. In fact, the Odd+11 method does not even require you to remember the 2-digit input year once you have started the mental calculation. All you have to remember is a single accumulator value in your head. For those familiar with computer architecture, this method can be implemented in a single-accumulator microprocessor with no need for extra registers or storage of variables in external memory.

Before we proceed, let us define the *7's complement* of a digit ***n*** as equal to ***7 - n.*** We will actually devote a whole section later to explain *7's complements* in more detail

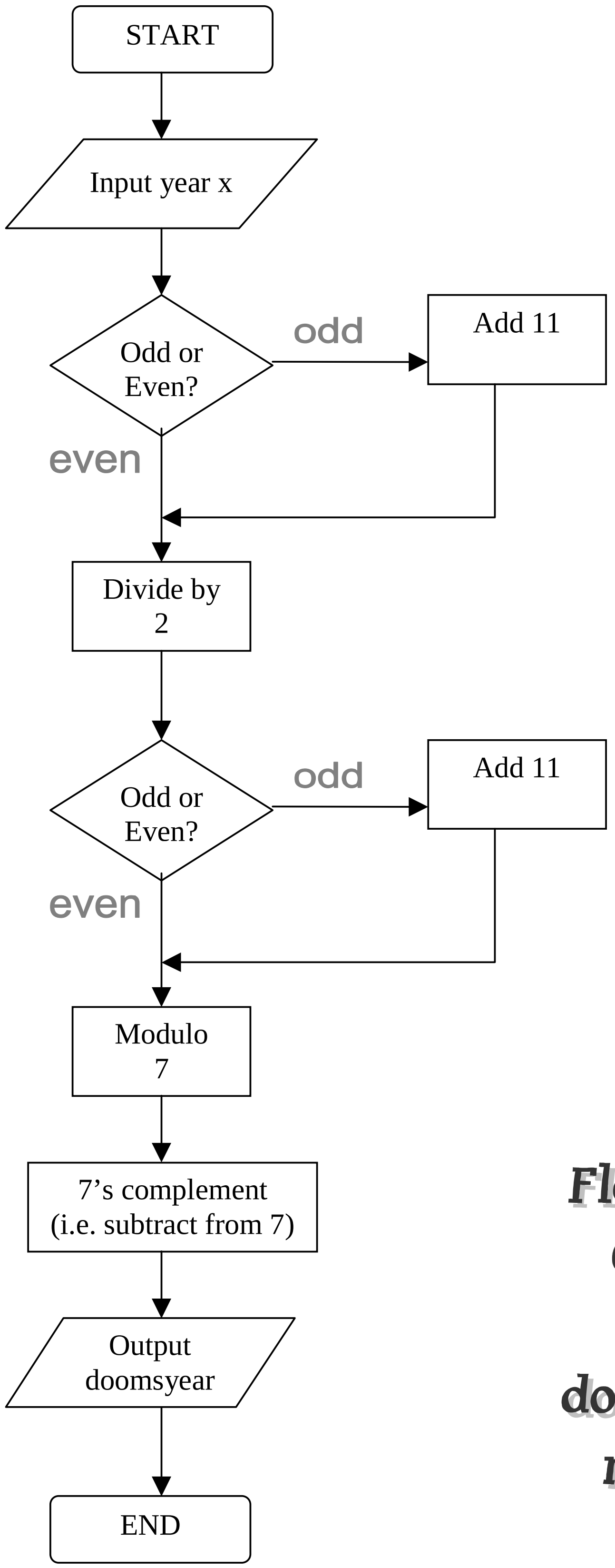

**Flowchart Of The Odd+11 doomsyear method**

**Examples:**

1) 1985: 85 is odd
→ plus eleven is 96
→ divided by two is 48 which is even
→ modulo seven is 6 because 48 - 42 = 6
→ whose seven's complement is **1**

2) 1999: 99 is odd
→ plus eleven is 110
→ divided by two is 55 which is odd
→ plus eleven is 66
→ modulo seven is 3 because 66 - 63 = 3
→ whose seven's complement is **4**

3) 1974: 74 is even
→ divided by two is 37 which is odd
→ plus eleven is 48
→ modulo seven is 6 because 48 - 42 = 6
→ whose seven's complement is **1**

4) 2040: 40 is even
→ divided by two is 20 which is even
→ modulo seven is 6 because 20 - 14 = 6
→ whose seven's complement is **1**

5) 2010: 10 is even
→ divided by two is 5 which is odd
→ plus eleven is 16
→ modulo seven is 2 because 16 - 14 = 2
→ whose seven's complement is **5**

6) 1988: 88 is even
→ divided by two is 44 which is even
→ modulo seven is 2 because 44 - 42 = 2
→ whose seven's complement is **5**

7) 2007: 07 is odd
→ plus eleven is 18
→ divided by two is 9 which is odd
→ plus eleven is 20
→ modulo seven is 6 because 20 - 14 = 6
→ whose seven's complement is **1**

8) 1998: 98 is even
→ divided by two is 49 which is odd
→ plus eleven is 60
→ modulo seven is 4 because 60 - 56 = 4
→ whose seven's complement is **3**

9) 1993: 93 is odd
→ plus eleven is 104
→ divided by two is 52 which is even
→ modulo seven is 3 because 52 - 49 = 3
→ whose seven's complement is **4**

10) 2000: 00 is even
→ divided by two is 00 which is even
→ modulo seven is 0 because 0 - 0 = 0
→ whose seven's complement is **0**

11) 2026: 26 is even
→ divided by two is 13 which is odd
→ plus eleven is 24
→ modulo seven is 3 because 24 – 21 = 3
→ whose seven's complement is **4**

12) 1935: 35 is odd
→ plus eleven is 46
→ divided by two is 23 which is odd
→ plus eleven is 34
→ modulo seven is 6 because 34 - 28 = 6
→ whose seven's complement is **1**

13) 2011: 11 is odd
→ plus eleven is 22
→ divided by two is 11 which is odd
→ plus eleven is 22
→ modulo seven is 1 because 22 - 21 = 1
→ whose seven's complement is **6**

Sidenote#1: We would like to mention one way of checking your calculations midway through the process. When performing the modulo 7 operator, the input should always be even. If it is not an even number, there must be an arithmetic error in your previous calculations .

Sidenote#2: The 7's complement of 0 is actually 7, but in modulo-7 arithmetic, seven is equivalent to zero.

**Seven's complement**

We previously defined the *7's complement* of a digit as that number subtracted from seven; i.e. the 7's complement of ***n*** is equal to ***7 - n***. In this section, we will provide a reason for using the concept of 7's complement in our algorithm and digress slightly about its inspiration from the computer arithmetic literature. Readers with no interest in computer arithmetic can safely skip the rest of the section, except perhaps to look at the table of values for 7's complement

| digit | 7's complement |
|---|---|
| 0 | 0 |
| 1 | 6 |
| 2 | 5 |
| 3 | 4 |
| 4 | 3 |
| 5 | 2 |
| 6 | 1 |

Table of values for 7's complement

We do not like negative numbers. Dealing with negative numbers make a process error-prone especially for mental calculations. Unfortunately, Conway's design of the doomsday algorithm necessitates subtraction from anchor dates. Sometimes, the result of this subtraction can be negative. Moreover, the design of the Odd+11 method also involves negation. This negation means that we want to do subtraction from other terms in the full evaluation of the doomsday algorithm.

One way to avoid negative numbers is to borrow a concept from the computer arithmetic literature called the *method of complements*[3]. Arithmetic circuit designers have discovered an elegant way of representing negative numbers in electronic circuits[4] by using the *1's complement* and *2's complement* of binary numbers. Likewise, they came up with analogous definitions for *9's complement* and 10*'s complement* of decimal digits. In general, for any base-M arithmetic system, they have defined the *M-1's complement* and *M's complement* of a number. We will not go into detail about the method of complements except to mention that our definition of the 7's complement of a single digit less than 7 is consistent with the definition of 7's complement given in the computer arithmetic literature[5] for radix-8 arithmetic of octal numbers.

The table of 7's complements provided above is easy to remember and calculate. We found out that in a lot of cases, calculating the 7's complement of a digit after performing a modulo 7 operation is easier than doing a negation modulo 7 in our head. By using the concept of 7's complement, we have avoided negative numbers altogether. Bob Goddard pioneered the use of 7's complement for the doomsday algorithm in his 2009 paper[6].

Mathematically speaking, we use the fact that calculating the modulo 7 of a negative number is the same as calculating the modulo 7 of the absolute value of that number first then getting the 7's complement of the result. In other words, for a positive integer x,

$$
\begin{aligned}
-x \ (\text{mod } 7) &= [-x \ (\text{mod } 7)] \ (\text{mod } 7) \\
&= [- (x \text{ mod } 7)] \ (\text{mod } 7) \\
&= 7 - (x \text{ mod } 7) \\
&= \text{seven's complement of } (x \text{ mod } 7)
\end{aligned}
$$

## The Group $Z_7$

We can use the language of abstract algebra to think about seven's complement. Consider the mathematical group $Z_7$ with addition modulo 7 as its binary operator. We have $Z_7=\{0,1,2,3,4,5,6\}$. The inverse of elements in $Z_7$ is calculated using seven's complement. That is, for all $x \in Z_7$ , the inverse of x is 7-x .

The days of the weeks form a group isomorphic to the group $Z_7$. Any time we do addition and subtraction of days in this group, we can just follow the arithmetic of $Z_7$. Likewise, every instance of negation can be replaced by a seven's complement operator. Conway's correspondence between the days of the week and $Z_7$ is given by the table:

| day of the week | $Z_7$ element |
|---|---|
| sunday | 0 |
| monday | 1 |
| tuesday | 2 |
| wednesday | 3 |
| thursday | 4 |
| friday | 5 |
| saturday | 6 |

Using this table, we can do arithmetic on the days of the week. For example:

1) Two days after **monday** = monday + 2 = 1 + 2 = 3 = **wednesday**
2) Three days after **tuesday** = tuesday + 3 = 2 + 3 = 5 = **friday**
3) Three days before **tueday** = tuesday - 3 = 2 - 3 = negative 1
   using seven's complement: negative 1 = 7 - 1 = 6 = **saturday**

4) Eighteen days after **friday** = friday + 18 = 5 + 18 = 23
   in modulo 7 arithmetic: 23 = 2 = **tuesday**

5) Eighteen days before **friday** = friday - 18 = 5 - 18 = negative 13
   in modulo 7 arithmetic: 13 = 6
   using seven's complement: negative 6 = 7 - 6 = 1 = **monday**

6) Sixty days before **sunday** = sunday - 60 = 0 - 60 = negative 60
   in modulo 7 arithmetic: 60 = 4
   using seven's complement: negative 4 = 7 - 4 = 3 = **wednesday**

**Tips and Tricks**

1) Plus Eleven

Adding eleven is very easy to perform mentally for 2-digit numbers. This is a byproduct of the fact that our number system is base-10. Adding eleven is as easy as incrementing both decimal digits of the input year. The only difficulty with adding eleven might occur with numbers that end with a 9. Our tip for handling such numbers mentally is to break 11 as 1+10. So, for input numbers that end with a 9, increment it first by 1 then add 10. For example, for input number 29, add one to get 30 then add ten to get 40. For input number 89, add one to get 90 then add ten to get 100.

2) Modulo Seven

Performing modulo 7 mentally is a bottleneck of the Odd+11 method. Our tip for performing it quickly and reliably is to bite the bullet and completely memorize all the multiples of 7 from 0 to 70. We all had to do this when we were learning the multiplication table in elementary school. It is worthwhile to refamiliarize ourselves and know the multiples of 7 by heart in order to perform modulo 7 quickly and reliably. Another tip to make this process easier to perform mentally is to use the concept of 7's complement. We can avoid error-prone carry-based subtractions by selecting the closest multiple of 7 to the input number within the decade; regardless of whether the multiple is greater than or less than the input number. If the multiple of 7 happens to be larger than the input number, we just perform 7's complement afterwards. Here is an outline of how to calculate modulo 7 with a 7's complement trick:

a) Find the multiple of 7 within the decade closest to the input number
b) Find the difference between the input number and the closest multiple of 7
c) If the multiple of 7 is greater than the input number, perform 7's complement on the result

Example:
i) Find 32 mod 7
   Closest multiple of seven is 35 → 32 - 35 = -3 → use seven's complement to get 4
ii) Find 58 mod 7
   Closest multiple of seven is 56 → 58 - 56 = 2 (no need for seven's complement)
iii) Find 40 mod 7
   Closest multiple of seven is 42 → 40 - 42 = 2 → use seven's complement to get 5
iv) Find 62 mod 7
   Closest multiple of seven is 63 → 62 - 63 = 1 → use seven's complement to get 6
v) Find 24 mod 7
   Closest multiple of seven is 21 → 24 - 21 = 3 (no need for seven's complement)

Note that a 7's complement operation is required at the tail end of the Odd+11 method. Hence, with this trick, one might need to perform two consecutive 7's complement operations back to back. The 7's complement of a number's 7's complement is , of course, the number itself. It is easy to modify this trick slightly to compute the negative modulo 7 of a number quickly.

## What's In a Name?

We named this method as the Odd+11 method because there are two branching points within the algorithm where one has to check whether the accumulator is odd or even. If the result is odd, one has to add 11 to the accumulator. We believe that naming it as such will make the method easier to remember.

## Doomsday Calculation Order

We previously stated that one of our design criterion for this method was to avoid the use of temporary variables to recall during the calculation. This method has a genuine benefit in making the evaluation of the full Doomsday algorithm easier if one calculates the doomsyear term first before the other terms. Let us elaborate.

Recall that the key equation for the Doomsday algorithm for input date MM/DD/CCYY as

day_of_the_week = ( doomscentury + doomsyear + doomsmonth ) mod 7

where:

doomscentury(CC) is a function of the input date's century
doomsyear(YY) is a function of the input date's 2-digit year within a century
doomsmonth(MM,DD) is a function of the input date's calendar month and day.

Our doomscentury term is actually the same as what Conway and others[5] define as the Doomsday (February 29) of a given centennial year. Our doomsmonth term is what takes advantage of the fact that April 4, June 6, August 8, October 10, and December 12 all fall on the same day of the week for any calendar year

Now, the evaluation of the three functions: doomscentury, doomsyear, and doomsmonth given above are order-independent. That is, one can calculate each of these terms out of order with no effect on the result. Actually, one has the freedom to choose which term to calculate first as he pleases. This is, of course, with the caveat that one has to remember the numerical result of that term to add with the other results in the end.

With that said, we propose that the calculation of the doomsyear term be done first in the full evaluation of the Doomsday algorithm because it requires no temporary variable to be remembered. That is, one has a clean slate by calculating the doomsyear term ahead of the doomscentury and doomsmonth term. This makes the calculation of the full Doomsday algorithm less error-prone. Here is our proposed calculation order for the terms of the Doomsday algorithm

1) Calculate doomsyear term first
2) Calculate doomscentury next; and add it to the already calculated doomsyear. In practical cases, doomscentury is simply memorized and recalled from memory
3) Calculate the value of doomsmonth and add it to the running-sum accumulated from the previous terms
4) Perform one last modulo 7 calculation to get the day of the week.

**Mathematical Proof**

We will now prove the correctness of the Odd+11 method.

*Theorem*: For x ∈ {0, 1, 2, … 99},

$$x+\left\lfloor \frac{x}{4} \right\rfloor \equiv - \left[\frac{x+11(x\,mod\,2)}{2} + 11(\frac{x+11(x\,mod\,2)}{2}\,mod\;2)\right] (mod\,7)$$

*Proof:* It suffices to show that

$$0 = x + \left\lfloor \frac{x}{4} \right\rfloor + \left[\frac{x+11(x\,mod\,2)}{2} + 11(\frac{x+11(x\,mod\,2)}{2}\,mod\;2)\right] (mod\,7)$$

or that

$$x + \left\lfloor \frac{x}{4} \right\rfloor + \left[\frac{x+11(x\,mod\,2)}{2} + 11(\frac{x+11(x\,mod\,2)}{2}\,mod\;2)\right]$$

is divisible by 7. Let us denote this as expression (1).

Now, let

$$k = \left\lfloor \frac{x}{4} \right\rfloor \quad , \quad r = x\;mod\,4$$

It follows that x = 4k + r. By substitution, we get expression (1) as

$$4k+r+k+\frac{4k+r+11(4k+r)\,mod2}{2}+11\left[\frac{4k+r+11(4k+r)\,mod\,2}{2}\,mod2\right]$$

$$= 7k + r + \frac{r+11(4k+r)\,mod\,2}{2} + 11\left[\frac{4k+r+11(4k+r)\,mod\,2}{2}\,mod2\right]$$

Now let us examine the each of the four possible cases for r:

<u>Case#1</u>: r = x mod 4 = 1

Expression (1) simplifies to

$$7k + 1 + \frac{1 + 11(4k+1)\,mod2}{2} + 11\left[\frac{4k + 1 + 11(4k+1)\,mod2}{2}\,mod\,2\right]$$

$$= 7k+1+\frac{1+11}{2}+11\left[\frac{4k+1+11}{2}\,mod\,2\right] \quad = \quad 7k+1+6+11\left[(2k+6)\,mod2\right]$$

$$= 7k + 1 + 6 + 0 \quad = \quad 7k + 7$$

which is divisible by 7.

<u>Case#2</u>: r = x mod 4 = 2

Expression (1) simplifies to

$$7k + 2 + \frac{2 + 11(4k+2)\,mod2}{2} + 11\left[\frac{4k + 2 + 11(4k+2)\,mod2}{2}\,mod\,2\right]$$

$$= 7k+2+\frac{2+0}{2}+11\left[\frac{4k+2+0}{2}\,mod\,2\right] \quad = \quad 7k+2+1+11\left[(2k+1)\,mod2\right]$$

$$= 7k + 3 + 11 \quad = \quad 7k + 14$$

which is divisible by 7.

Case#3:   r = x mod 4 = 3

Expression (1) simplifies to

$$7k + 3 + \frac{3 + 11(4k+3)\,mod 2}{2} + 11\left[\frac{4k + 3 + 11(4k+3)\,mod 2}{2}\,mod\,2\right]$$

$$= 7k+3+\frac{3+11}{2}+11\left[\frac{4k+3+11}{2}\,mod\,2\right] \quad = \quad 7k+3+7+11\,[(2k+7)\,mod 2]$$

$$= 7k + 3 + 7 + 11 \quad = \quad 7k + 21$$

which is divisible by 7.

Case#4:   r = x mod 4 = 0

Expression (1) simplifies to

$$7k + \frac{11(4k)\,mod 2}{2} + 11\left[\frac{4k + 11(4k)\,mod\,2}{2}\,mod\,2\right]$$

$$= 7k + 0 + 11\left[\frac{4k}{2}\,mod\,2\right] \quad = \quad 7k + 0 + 0 \quad = \quad 7k$$

which is divisible by 7.

Hence, we have shown that expression (1) is divisible by 7 for all 4 cases of the possible values of x mod 4.  QED

## Conclusion

We presented an alternative acceleration method for calculating the doomsyear term of the Doomsday algorithm. This method is the only method we know of that does not require a divisibility-by-4 test as part of the algorithm. On the average, we have found this method to be easier to calculate than the Decade method described in part 1 of this paper. In practice, we use both methods to check and verify each other's results.

### Acknowledgements

The authors would like to thank YingKing Yu for insightful discussions of the Doomsday algorithm. YingKing reminded us of the connection between 7's complement and negation. He also provided a new mnemonic for January and February on leap years, 1/11 and 2/22. The authors would also like to thank Bob Goddard and Lawrence Baker for proofreading comments and suggestions.

## Appendix: Mike Walters' Original Doomsyear Method

Mike Walters' method for calculating doomsyear can be summarized by the following equation:

$$doomsyear(x) = -\frac{x + 11(x\,mod 4)}{2} \quad (mod\,7)$$

Mike Walters actually presented this method in a different way. He did not provide an explicit formula and did not calculate *11(x mod 4)* directly. Instead, he iteratively adds multiples of 11 to x until the sum is divisible by 4.

Let us prove the correctness of Mike Walter's method:

**Theorem**: For x ∈ {0, 1, 2, … 99},

$$x + \left\lfloor \frac{x}{4} \right\rfloor \equiv -\frac{x + 11(x\,mod\,4)}{2} \quad (mod\,7)$$

Proof: It suffices to show that

$$0 = \left( x + \left\lfloor \frac{x}{4} \right\rfloor + \frac{x + 11(x\,mod\,4)}{2} \right) \quad (mod 7)$$

or that

$$x + \left\lfloor \frac{x}{4} \right\rfloor + \frac{x + 11(x\,mod 4)}{2}$$

is divisible by 7.

Since $\left\lfloor \frac{x}{4} \right\rfloor = \frac{x - (x\,mod 4)}{4}$,

$$x + \left\lfloor \frac{x}{4} \right\rfloor + \frac{x + 11(x\,mod 4)}{2} = x + \frac{x - (x\,mod 4)}{4} + \frac{x + 11(x\,mod\,4)}{2}$$

$$= \frac{4x + x - x\,mod 4 + 2x + 22(x\,mod 4)}{4} = \frac{7x + 21(x\,mod\,4)}{4}$$

$$= 7\frac{x + 3(x\,mod\,4)}{4}$$

which is, indeed, divisible by 7. QED